# A multimodal computer-vision based pipeline for eye-hand coordination quantification during the Nine-Hole Peg Test

Rajvardhan Gadde[1,2 *], Mahya Beheshti[1,2*], Anjali Rajkumar[1], Silvana L. Costa[4], John-Ross Rizzo[1,2,3], Todd E. Hudson[1]

## Abstract

The Nine-Hole Peg Test (9-HPT) is a widely used clinical measure of manual dexterity in multiple sclerosis (MS), but its primary outcome—task completion time—provides limited information about the visuomotor mechanisms contributing to impaired performance. Instrumented approaches can provide richer kinematic information but often require modifications to the task or physical instrumentation that may alter natural hand-object interaction. We developed a multimodal computer-vision pipeline that enables synchronized analysis of gaze, hand kinematics, and object state during an unmodified 9-HPT. Gaze was recorded using Pupil Labs Invisible glasses, three-dimensional hand landmarks were reconstructed using WiLoR, and a custom YOLO26x-seg model detected and tracked the pegboard, holes, pegs, and collection bin. Eye-tracking, hand, and object data were temporally synchronized at the video-frame level and spatially registered to a common pegboard-centered coordinate system using frame-specific projective transformations. The pipeline further integrated hole-state transitions, hand-zone classification, and fingertip kinematics to identify PICK and PLACE events and segment hand transport and stationary phases, while fixation and saccade labels were derived from the eye-tracking system. Automated detections were visually reviewed and corrected when necessary. The resulting framework provides synchronized, event-resolved measures of gaze behavior, hand movement, object interaction, and eye-hand coordination while preserving standardized 9-HPT administration. This approach extends the 9-HPT beyond completion time and provides a reproducible platform for investigating the visuomotor mechanisms underlying dexterity impairment in MS and other neurological disorders.

[1] Department of Rehabilitation Medicine, NYU Langone Health, 244 E38th St., New York, NY, 10016, USA.
[2] Department of Mechanical & Aerospace Engineering, NYU Tandon School of Engineering, New York, NY, 10010, USA.
[3] Department of Neurology, NYU Langone Health, 222 E40th St., New York, NY, 10016, USA
[4] Neuropsychology & Neuroscience Laboratory, Kessler Foundation, West Orange, NJ, USA
*Authors contributed equally

## 1 Background

Multiple sclerosis (MS) is a chronic neurodegenerative disease in which upper extremity dysfunction is one of the most common and disabling manifestations, frequently impairing manual dexterity and reducing independence in activities of daily living. Accurate assessment of upper extremity function is therefore essential for monitoring disease progression, evaluating therapeutic interventions, and measuring clinical outcomes. The Nine-Hole Peg Test (9-HPT) is a standardized assessment of upper extremity fine motor dexterity incorporated into the Multiple Sclerosis Functional Composite (MSFC) due to its sensitivity to disease progression, ease of administration, and widespread clinical acceptance [1-3]

Although the 9-HPT is considered the clinical gold standard for assessing manual dexterity in individuals with MS, its primary outcome measure is task completion time. While completion time provides a global measure of performance, it provides limited insight into the specific motor, sensory, or visuomotor impairments underlying reduced performance [4, 5]. For example, prolonged completion time may result from tremor, impaired grasp precision, inefficient hand transport, visual processing deficits, or impaired eye-hand coordination. Because these impairments have distinct clinical implications and may require different rehabilitation strategies, completion time alone cannot fully characterize the mechanisms contributing to impaired dexterity.

Technology-assisted approaches have been developed to provide greater insight into movement characteristics during pegboard tasks. Instrumented systems, such as the Virtual Peg Insertion Test, have demonstrated that kinematic and force-based measures can reveal movement abnormalities that are not captured by completion time alone [6]. More recently, computer vision and machine learning approaches have enabled markerless assessment of the 9-HPT by estimating hand kinematics and task performance directly from video recordings [7, 8]. However, these approaches have primarily focused on quantifying hand movements, without capturing how visual attention is coordinated with these motor actions during object manipulation.

Eye-hand coordination is fundamental to visually guided object manipulation, with gaze typically preceding and guiding subsequent hand movements during reaching, grasping, and object placement [9]. The 9-HPT requires continuous integration of visual information, hand movements, and interactions with task objects, yet current assessments cannot capture the dynamic coordination between these components during standardized clinical testing. A synchronized multimodal approach that integrates these complementary measures would therefore provide greater insight into the visuomotor strategies underlying dexterity performance while preserving the established 9-HPT protocol.

In this study, we present a multimodal analysis pipeline for synchronized assessment of eye-hand coordination during standard administration of the 9-HPT. The methodology integrates eye tracking, markerless hand tracking, and object tracking within a unified framework to automatically identify task events and quantify the dynamic interactions between gaze behavior, hand kinematics, and object manipulation throughout performance. By preserving the standardized 9-HPT protocol while expanding the information extracted from each trial beyond completion time and isolated movement measures, this reproducible pipeline provides a practical approach for investigating the mechanisms underlying dexterity performance in multiple sclerosis and other neurological disorders

## 2 Methods

### 2.1 Apparatus and Equipment

The experimental setup consisted of three integrated systems for multimodal data acquisition: a hand tracking system, an object detection and segmentation system, and an eye tracking system.

**2.1.1 Hand Tracking.** Three-dimensional hand landmarks were reconstructed using WiLoR, a transformer-based hand reconstruction model. Hand tracking was performed on frames acquired from the Pupil Labs Invisible scene camera at 25 frames/s. When WiLoR failed to detect the participant's hand, YOLO-derived hand bounding boxes were used to define a higher-resolution region of interest for a second WiLoR inference pass.[10]

**2.1.2 Eye Tracking.** Gaze position was recorded using Pupil Labs Invisible Glasses [11] sampled at 200 Hz. Prior to each session, the eye tracker was offset corrected by looking at the stable target and correcting the gaze pointer accordingly.

**2.1.3 Object Detection and Segmentation.** A YOLO26x segmentation model [12, 13] was trained on a 2151-image augmented custom dataset to detect and classify task-relevant objects, including individual pegs, holes in the pegboard, the collection bin, and the pegboard itself. Detected objects were tracked across frames using the BoT-SORT [14] multi-object tracking algorithm to preserve object identity through brief occlusions.

The training ran for 300 epochs with 16 batch size done on Nvidia A100 GPU. Data augmentation comprised mosaic, horizontal flip, HSV jitter, random scale (±50%) and translation (±10%). The checkpoint with highest validation fitness was used.

### 2.1.4 Nine-Hole Peg Test Procedure

The Nine-Hole Peg Test apparatus consists of a shallow bin holding nine small pegs adjacent to a 3×3 grid of holes [1]. Participants were instructed to place the pegs into the holes one at a time as quickly as possible, then remove them individually and return them to the bin.

Participants were seated at a table with the pegboard positioned at a comfortable reaching distance, after which the eye tracker was offset corrected. Participants completed the 9-HPT with their dominant hand first, followed by their non-dominant hand. In accordance with the standardized protocol, two trials were performed per hand, and the average completion time was recorded [2]. Throughout each trial, hand position, gaze position, and object detection data were recorded continuously and synchronized via timestamps.

## 2.2 Data Processing

### 2.2.1 Hand Kinematic Processing

Hand position was estimated using WiLoR [10]. The handedness of the trial was identified based on whichever hand is observed to have predominant movement, while detections corresponding to the experimenter's hands or inactive hands were excluded.

When WiLoR failed to detect the participant's hand, YOLO-generated hand bounding boxes were used to define regions of interest (ROIs), within which hand detection was repeated at an increased image resolution.

Hand trajectories were subsequently corrected for scene-camera motion using the head-motion compensation procedure described in Section 2.2.2.

### 2.2.2 Head-Motion Compensation

To compensate for participant head motion, a Lucas–Kanade sparse optical flow tracker [15] estimated camera displacement from a user-defined stable background region. Up to 200 Shi-Tomasi corner features were tracked frame-to-frame with a forward-backward consistency and the median displacement across points (median rather than mean, for robustness to individual mistracked points) was taken as the frame's motion estimate; features were refreshed every 60 frames or sooner if fewer than 30 remained. The resulting displacement series was smoothed with a Savitzky-Golay filter and subtracted from all landmark and gaze coordinate columns post hoc.

### 2.2.3 Eye Tracking Processing

Gaze data from the Pupil labs Invisible glasses were exported as tab-separated files sampled at 200 Hz, interleaved in the same export with gyroscope channels. Two complementary eye-related signals were extracted: 1) gaze event labels (fixation/saccade) from Pupil cloud and 2) spatial gaze coordinates from the scene video. These signals were used separately for temporal classification and spatial registration, respectively.

### 2.2.4 Eye Movement Classification

Eye-movement classification (fixation vs. saccade) was taken directly from the Pupil cloud itself rather than being derived by the vision pipeline: fixations and saccades were classified by Pupil labs, and this classification is exported as an "Eye movement type" field in the raw CSV. Each recording was first manually synchronized and trimmed to the task video using a shared cut point identified by a reviewer; within the resulting file, the trial of interest was then extracted automatically using explicit start/end event markers embedded in the CSV's Event column, or, for files that had already been trimmed to a single trial upstream, using the first and last eye-tracker sample timestamps in the file directly. Each 200 Hz eye-tracker sample within the trial window was mapped onto a video frame index by its recording timestamp between the trial's start and end timestamps across the total number of video frames. Because several eye-tracker samples typically fall within the duration of a single video frame (approximately eight, at 200 Hz eye-tracking against a 25-frame/s scene video), the majority eye-movement label among the samples mapping to a given frame was assigned as that frame's label. This per-frame Fixation/Saccade label is the signal used directly for event annotation (see Section 2.4.3): saccade-labeled frames were treated as the ocular transport phase and fixation-labeled frames as the ocular stationary phase.

### 2.2.5 Object Detection and Tracking

Pegboard objects, empty and peg-filled holes, the peg bin, the board frame, and each hand were detected every frame with a custom-trained YOLO model, run only on a manually defined region of interest (ROI) enclosing the pegboard rather than the full frame; detected boxes were offset back into full-frame coordinates before being combined with the hand-tracking and gaze signals. Detection was performed jointly with multi-object tracking (BoT-SORT) rather than independent per-frame detection, so that individual holes, the bin, and each hand retained a consistent identity across the trial through brief occlusions. The tracker itself was queried with a permissive confidence floor (0.01) and IoU threshold (0.1); precision was instead controlled by a second, per-class confidence threshold applied after tracking (0.5 for the board frame; 0.1 for all other classes), allowing classes prone to false positives to be tuned independently without discarding lower-confidence board detections needed to anchor the scene. When multiple partial board detections occurred in a single frame, they were merged into a single consolidated board bounding box, extended out to a fixed set of stable positional landmarks corresponding to the board's holes, so that the board region remained consistent even when the detector fragmented the board into pieces.

## 2.3 Trial state classification

### 2.3.1 Hole-State Tracking and Hand-Occlusion Handling

Two hole-related classes were distinguished by the detector, empty and peg-filled holes. Per-frame count of each is the raw signal from which peg placement and removal is ultimately derived. Object detection alone was not sufficient to reliably count peg-filled holes, however, because a hand or wrist frequently occludes one or more holes while a peg is being placed or removed, causing the detector to transiently under or mis-count mid-manipulation. To address this limitation, two complementary hand-location–gated mechanisms were implemented.

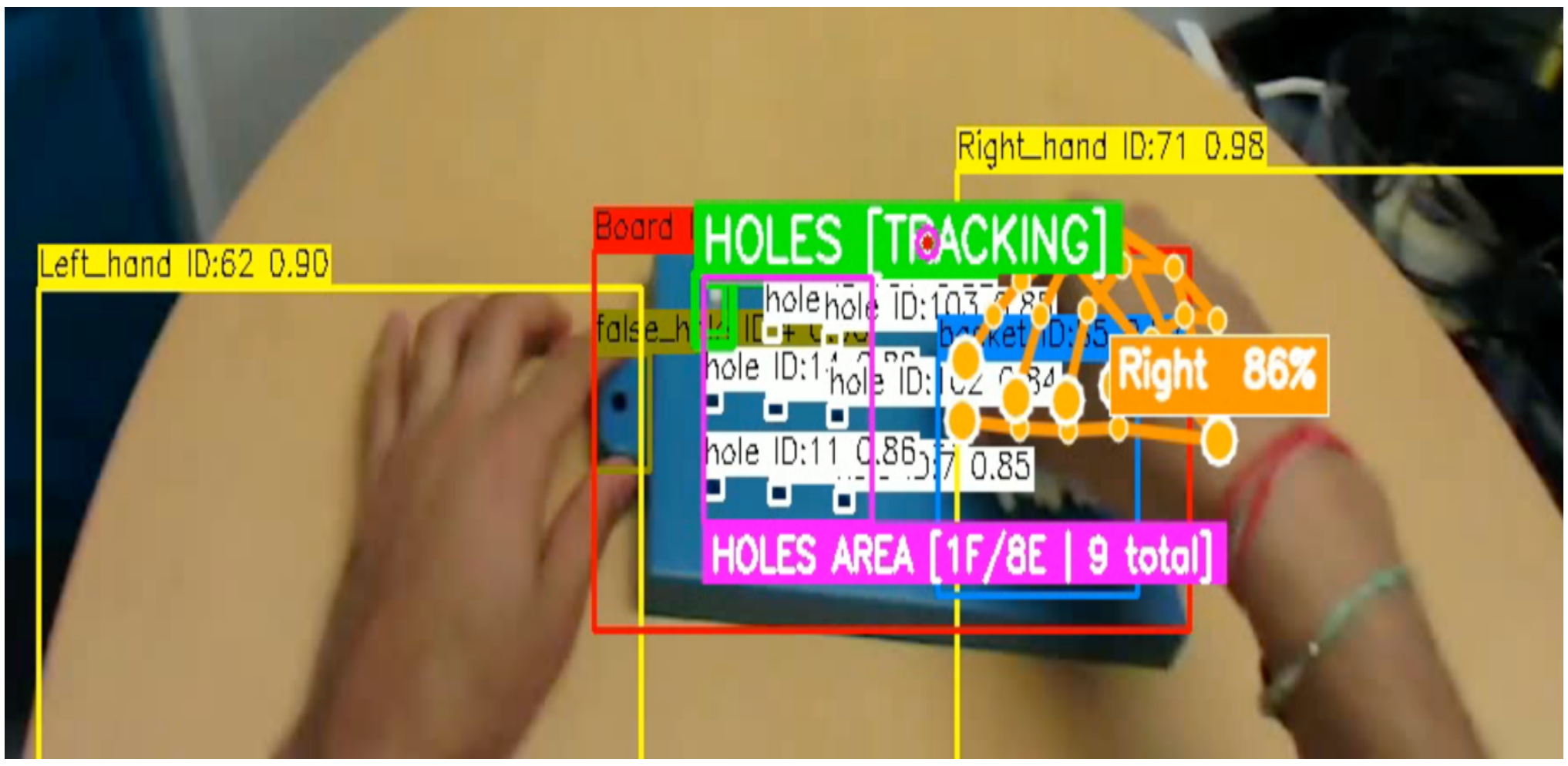

**Figure 1.** Example frame showing markerless hand tracking, pegboard segmentation, and hole-state detection during the 9-HPT

**2.3.2 Individual hole-level gating.** At the level of an individual hole, a fixed hole registry was established once at least seven of the board's nine holes had been observed simultaneously in a single frame; each hole was thereafter tracked across frames by nearest-neighbor matching on position (60 px tolerance). A hole was flagged as occluded, and its fill-state held at its last known value rather than updated whenever the overlap between its expected disk region and any tracked hand's bounding box exceeded 0.15; the state change (empty to filled) was only confirmed once the hole became visible again.

**2.3.3 Board-level hand gating.** At the level of the whole board region, whenever any tracked hand was located within the bounding region enclosing all the holes (Fig. 1), the pipeline suspends real-time updates to the overall peg-filled holes count entirely, carrying forward the last count observed immediately before the hand entered rather than trusting the detector's live, occlusion-degraded count for that frame. The fill-state of every hole was frozen at the moment of hand entry; only once the hand had remained continuously outside the region for two consecutive frames  was the post-exit hole state compared against the pre-entry snapshot, with any net difference maked as a peg-filled holes or peg-removed event at that point. In effect, hole-state changes were accepted only once the obstructing hand had fully withdrawn, rather than being read continuously during manipulation itself. This hand-location gating is what prevented false fill/empty flicker from hand occlusion during active placement.

**2.3.4 Offline Hole-State Smoothing for Event Detection**

In parallel with this gated, real-time count, the raw (ungated) per-frame peg-filled holes count was retained separately and smoothed offline with a 7-frame median filter; stable plateaus of at least 10 consecutive frames at a constant count were used to derive a second, offline set of hole-state transition events that do not lag behind the true placement moment the way the hand-gated real-time count does. It is this offline, smoothed signal, not the real-time, hand-gated count that anchors PLACE events in the automatic event detector described below (see Section 2.4.2). Thus, the hand-gated approach was used to maintain a stable estimate of hole state during hand occlusion, whereas the offline smoothed signal was used to determine the timing of peg-placement events.

### 2.3.5 Hand-Zone Classification

Independently of hole-state tracking, each hand's coarse location inside the bin region, on/near the board, or in transit between the two  was classified every frame from the positions of five fingertip landmarks relative to the bin and board bounding boxes (30 px and 20 px margins, respectively). Zone-entry and zone-exit transitions were logged as discrete events and used both for movement-phase segmentation and to define the bin-visit window within which PICK events were subsequently searched (see Section 2.4.2).

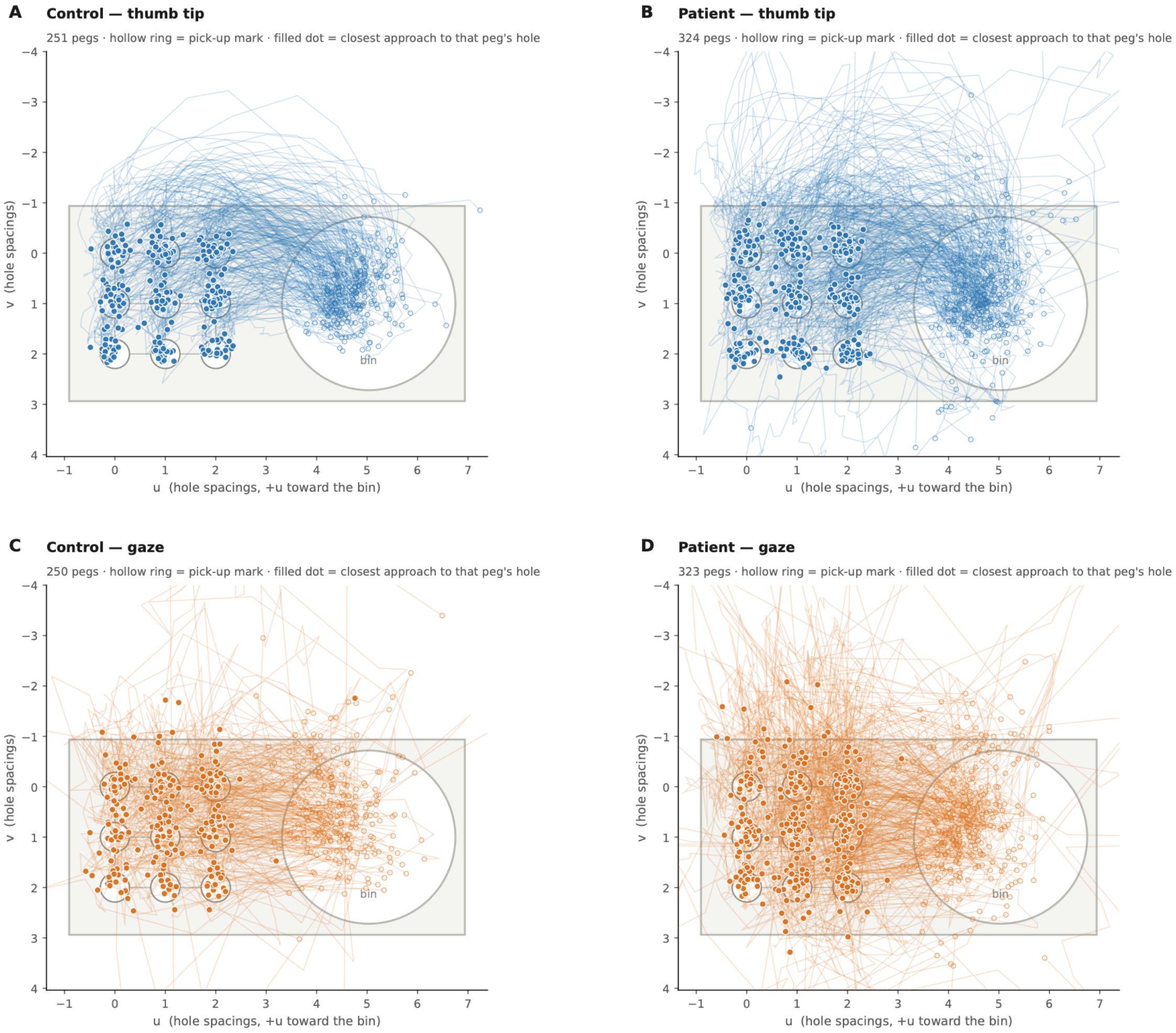

**Figure 2.** Sample data **(a), (b)** showing the thumb tip position on the board

**(c), (d)** showing the gaze position on the board.

### 2.3.6 Spatial Registration to Pegboard Coordinates

Pixel distances in the head-mounted scene-camera image could not be directly converted to physical distance because the apparent spacing between holes varied with viewing perspective, participant head position, and distance from the pegboard. Although the physical center-to-center distance between adjacent holes was fixed at 32 mm, this spacing differed in pixels across the image. Therefore, a projective transformation was used to map image-based gaze and hand coordinates into a common pegboard-centered coordinate system.

The pegboard was represented as a 3 × 3 lattice with coordinates ranging from (0,0) to (2,2), where u increased from left to right and v increased from the far row to the near row. This coordinate system provided a standardized reference frame for comparing gaze and thumb-tip positions across frames and participants.

Hole reference points were derived from detections of both empty and peg-filled holes. Because insertion of a peg increased the height of the detection bounding box and shifted its centroid upward, the midpoint of the bottom edge of each hole bounding box was used as the initial spatial reference point. These reference points were then used to estimate the 3 × 3 pegboard lattice.

A frame-specific homography was estimated to transform scene-camera coordinates into pegboard coordinates. The method used to estimate the lattice depended on the number of visible holes. When all nine empty or peg-filled holes were visible, neighboring hole positions were used to assign each detection to its corresponding row and column, and a homography was estimated directly. When six to eight holes were visible, the lattice from the nearest successfully solved frame was propagated using head-motion displacement estimated by the optical-flow tracker and then refitted to the available detections. A homography was used when eight holes were visible, whereas an affine transformation was used when six or seven holes were visible.

When fewer than six holes were visible, typically because of hand occlusion during peg placement, the four corners of the lattice from the nearest successfully solved frame were propagated according to the estimated head motion. A homography based on these propagated corners was then used to estimate the positions of all nine holes. When at least two holes or peg-filled holes remained visible, the estimated lattice was further aligned to the observed detections.

The resulting frame-specific homography was applied to gaze and thumb-tip coordinates, transforming both signals from scene-camera pixels into the common pegboard coordinate system. Because the lattice was initially defined using the bottom-center points of the hole detections, these locations were subsequently shifted to represent the physical centers of the 10-mm-diameter holes.

Finally, distances expressed in pegboard coordinates were converted to millimeters using the known 32-mm center-to-center spacing between adjacent holes. This transformation allowed gaze and thumb positions, as well as their distances from target holes, to be expressed in physical units and compared consistently across hole locations, viewing perspectives, and frames.

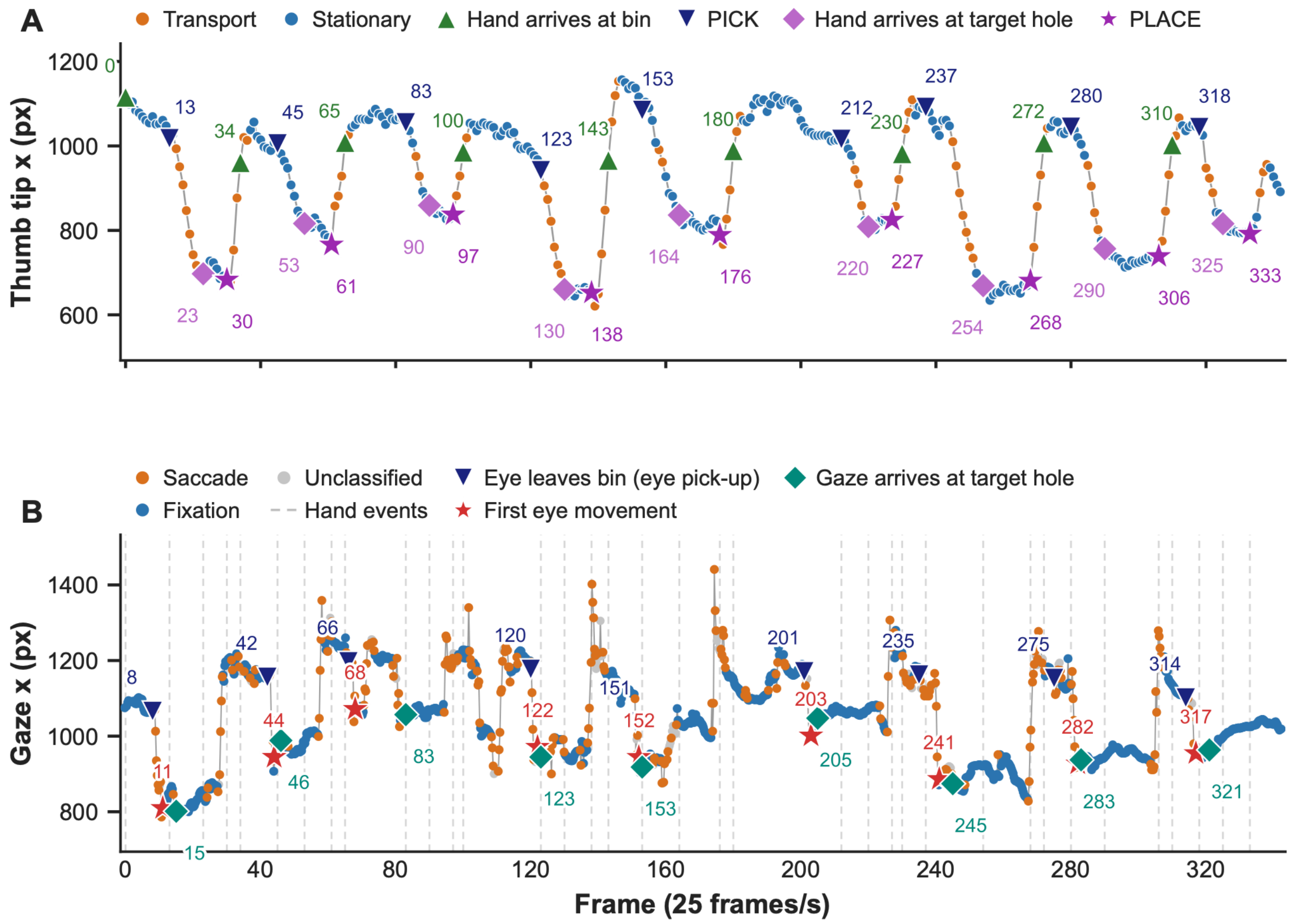

**Figure 3.** Hand and eye traces from one example trial (control participant, non-dominant hand), all nine peg placements (frames 0–342 at 25 frames/s).

**(a)** **Top:** horizontal thumb-tip position (px); orange = transport, blue = stationary (section 2.4.1). ▲ hand arrives at bin, ▼ PICK, ◆ hand arrives at target hole, ★ PLACE; the numbers are frame indices.

**(b)** **Bottom:** horizontal gaze position (px) at 100 Hz; (▼ eye leaves the bin (last bin fixation), ★ first eye movement toward the target, ◆ gaze arrives at the target hole. Light dashed lines repeat the hand events.

## 2.4 Event Detection & Annotation

### 2.4.1 Hand Movement Phase Classification

To classify movement phases during the 9-HPT into transport versus stationary periods, the horizontal position of the index fingertip was used. The absolute frame-to-frame x-displacement of the fingertip was smoothed using a 5-frame uniform filter, and frames with smoothed displacement greater than 30 pixels/frame were classified as transport, whereas frames with displacement at or below 30 pixels/frame were classified as stationary (Fig. 3).

### 2.4.2 Automated PICK and PLACE Event Detection

PICK and PLACE events were identified by integrating hole-state transitions, hand-zone information, and fingertip movement velocity. PLACE events were first identified from changes in hole occupancy, whereas PICK events were identified from the corresponding preceding hand interaction with the peg bin.

PLACE events were anchored to the offline, median-smoothed hole-fill signal described in Section 2.3.4 (rather than real-time, hand-gated count, which lags during manipulation): for each hole-fill transition, the algorithm searched backward up to 80 frames for the most recent stationary-phase velocity trough and treated that frame as the true placement moment. One PICK was then identified per PLACE as the most recent velocity trough that occurred while the hand was in the bin zone (Section 2.3.5), within the 150 frames preceding that PLACE. An additional event, PICK_INITIATION, was logged for each pick to mark when the hand came to rest in the bin: scanning backward from the PICK, it is the first frame after the smoothed displacement last exceeded an adaptive threshold (midway between the mean displacement around PICK events and the mean mid-transport displacement)

An alternate, valley-based event detector using Savitzky-Golay-filtered velocity was also implemented and run in parallel for cross-checking. This detector located velocity minima directly via peak detection on the inverted, pre smoothed speed signal (enforcing a minimum of 0.5 s between distinct events), cross-referenced each candidate against the sign of the horizontal velocity (a direction reversal is expected at a genuine pick or place) and against which zone (bin/board/transit) the hand occupied at that moment, and enforced strict alternation between PICK and PLACE across a trial. This detector additionally flagged brief (roughly 80 ms –1 s) gaps in hand-position tracking that occurred while the hand was in the bin zone as probable implicit grasps this is a case that arises when fast grasping motion briefly defeats the hand tracker by using the midpoint of the gap as the event frame. Agreement between this alternate detector and the primary transport/stationary detector above was used as a secondary check on event timing rather than as the primary detection method.

### 2.4.3 Eye Movement Event Identification

For the eye-movement traces, the frame-level Fixation/Saccade label described in Section 2.2.4 from the Pupil Labs Invisible eye tracker's own eye movement classification were used. Saccade-labeled frames were treated as the ocular transport phase and fixation-labeled frames as the ocular stationary phase, and the first eye movement toward each peg's target hole was identified from this signal for use in the eye-hand latency analyses.

### 2.4.4 Manual Review and Annotation Correction

All automatically detected events were visually reviewed against the source video before inclusion in downstream analyses. For trials where automatic detection failed or produced an implausible event (most commonly during atypical or interrupted movements), two complementary interactive tools were used to correct the record manually. A frame-by-frame video review tool allowed a rater to swap mislabeled hand identities, delete false or duplicate detections, manually place hand-landmark positions by clicking the wrist, index fingertip, and thumb tip when the hand tracker failed to detect a hand at all, and manually correct the hole fill-count when object detection under- or over-counted; a spatial grouping tool additionally allowed ambiguous hand-to-hole assignments to be resolved relative to a reviewer-chosen reference hole across a range of frames. A second tool, presenting synchronized hand- and eye-position traces frame by frame, was used to manually mark the first eye movement toward each of the nine target holes per trial (snapping to the nearest frame on click) and to locate the last stable fixation preceding the saccade into each such event, using a backward scan with a speed-baseline fallback when no unambiguous prior fixation existed. Corrections made with either tool propagated automatically into all downstream computations, so that manually annotated frames were treated identically to automatically detected ones in the final analysis.

## 3 Limitations

This pipeline classifies eye movements into saccades and fixation; more detailed event-related analyses require manual annotation, particularly for edge case or ambiguous events. Although gaze was sampled at 200 Hz, the scene camera used for markerless hand and object tracking operated at approximately 25 frames/s, limiting the temporal resolution of the video-derived kinematic measures. Future implementations could use a higher-frame-rate like head-mounted GoPro camera, to increase video sampling to approximately 200 frames/s, with gaze data mapped onto the higher-frame-rate video.